\documentclass[preprint,amsmath,amssymb]{revtex4}

\usepackage{graphicx}
\usepackage{dcolumn}
\usepackage{bm}

\begin{document}

\title{Warp Drive: A New Approach}

\author{Richard K Obousy}
\author{Gerald Cleaver}

\email{Richard\_K\_Obousy@baylor.edu}
\email{Gerald\_Cleaver@baylor.edu}
\affiliation{Baylor University, Waco, Texas, 76706, USA}

\date{\today}

\begin{abstract}
Certain classes of higher dimensional models suggest that the Casimir effect is a candidate for the cosmological constant. In this paper we demonstrate that a sufficiently advanced civilization could, in principal, manipulate the radius of the extra dimension to locally adjust the value of the cosmological constant. This adjustment could be tuned to generate an expansion/contraction of spacetime around a spacecraft creating an exotic form of field-propulsion.  Due to the fact that spacetime expansion itself is not restricted by relativity, a faster-than-light `warp drive' could be created. Calculations of the energy requirements of such a drive are performed and an `ultimate' speed limit, based on the Planckian limits on the size of the extra dimensions is found.\end{abstract}

\maketitle

\section{Introduction}

$\indent$A 1994 paper written by M. Alcubierre [2] demonstrated that, within the framework of general relativity, a modification of spacetime could be created that would allow a spacecraft to travel with arbitrarily large speeds. In a manner identical to the inflationary stage of the universe, the spacecraft would have a relative speed, defined as change of proper spatial distance over proper spatial time, faster than the speed of light. Since the original paper, numerous authors have investigated and built on the original work of Alcubierre.
\\
$\indent$Warp drives provide an unique and inspiring opportunity to ask the question `what constraints do the laws of physics place on the abilities of an arbitrarily advanced civilization'[22]. They also serve as a tool for teaching general relativity, as well as an exciting way to attract young students to the field by stimulating their imagination. In this paper an original mechanism to generate the necessary `warp bubble' is proposed. The main focus of the paper is to demonstrate that the manipulation of the radius of one, or more, of the extra dimensions found in higher dimensional quantum gravity theories, especially those that are based on or inspired by string/M-theory, creates a $\textit{local}$ asymmetry in the cosmological constant which could be used to propel a space vehicle.  
\\
$\indent$Warp drives have not been the sole interest of theoretical physicists as was demonstrated by the formation of the NASA Breakthrough Propulsion Program and the British Aerospace Project Greenglow, both of whose purpose was to investigate and expand on these ideas regarding exotic field propulsion.
\\
$\indent$At such an early stage in the theoretical development of the ideas presented in this paper, it is challenging to make predictions on how this `warp drive' might function. Naively one could envision a spacecraft with an exotic power generator that could create the necessary energies to locally manipulate the extra dimension(s). In this way, an advanced spacecraft would expand/contract the compactified spacetime around it, thereby creating the propulsion effect.
\\
$\indent$The first four sections of this paper review the necessary physics required to appreciate the new warp drive model. The remainder of the paper will introduce the propulsion concept. Calculations regarding speed limits and energy requirements will also be presented.
\section{The Cosmological Constant and Quantum Field Theory}
$\indent$Current observations of distant supernova [13] indicate that the universe is expanding. This expansion is realized in Einstein's equations   .

\begin{equation}R_{\mu\nu} -\frac{1}{2}Rg_{\mu\nu}=8\pi G T_{\mu\nu}-\Lambda g_{\mu\nu}     \end{equation}

through the cosmological constant term, $\Lambda$ This `Lambda' term, as it is known, provides a necessary addition to the gravitational field to correlate with what is observed in nature.
\\
$\indent$The fundamental origin of Lambda is still a mystery nearly a century after its introduction into cosmology. Physicists are not certain $\textit{what}$ generates the cosmological constant term in Einstein's equations; we simply know that it is there. Several ideas exist as to the nature of this field. Dark energy, for example, is a popular contemporary phrase for the Lambda term. Efforts have been made to explain the cosmological constant using the more modern quantum field theory, created after Einstein and his gravitational equations [7].
\\
$\indent$Quantum field theory (QFT) is widely regarded as one of the most, if not $\textit{the}$ most, successful physical theories of all time. Its predictions have been verified in particle accelerators; no experiment has ever shown a result that contradicts QFT. It would thus seem only natural to try to account for the cosmological constant using QFT. The calculations [8] are based on summing all the zero point oscillations with a Planck scale cut-off which would give us an estimate to the overall vacuum energy density $E_{vac}$. 
\begin{equation}
\left< E_{vac}\right> =\frac{1}{2}\int_0^M \frac{d^3k}{(2 \pi)^3}\sqrt{k^2+m^2} \approx \frac{M^4}{16\pi^2}
\end{equation}
$\indent$The energy density predicted using this equation is of order $10^{71} (GeV)^4$, which conflicts with the observed value of $10^{-48} (GeV)^4$. Indeed, this is genererally viewed as, by far, the worst predication of theoretical physics-being off by a factor of $10^{119}$! This failure of quantum theory has recently been reexamined using brane world scenarios born of string theory [11], [5], [14], [1], [10], [24].
\\
$\indent$One (partial) fix to the vacuum energy calculations is the introduction of supersymmetry (SUSY). The basic idea is that all known particles have an associated superparticle whose spin differs by exactly one half (in $\hbar$ units). In the case of unbroken SUSY, i.e., when particle and supersymmetric partner have exactly equal masses, the superparticle additions to the vacuum energy perfectly cancel the particle contributions and the resulting vacuum energy is reduced to zero. However, when SUSY is broken and the difference of the particle mass and the supersymmetric partner mass is on a scale of $M_S \approx 10^4 GeV$, then the resulting vacuum energy is on the order  $(M_S)^4 \approx 10^{16} GeV$ and the discrepancy is (only) off by a factor of $10^{60}$! In this paper we assume that even the SUSY-breaking contribution is somehow cancelled and that all contributions to the vacuum energy density are higher dimensional.
\section{Kaluza Klein Models}
$\indent$Kaluza-Klein (KK) theory was an early attempt to unify gravitation with electromagnetism. Kaluza initially postulated that a fifth spatial dimension could be introduced into Einstein's equations [16]. When the equations were solved the additional dimension would generate not only the standard metric, but also a vector field $A_\mu$  that could be associated with the electromagnetic field and a scalar field $\phi$. Kaluza's starting point was the metric:
\begin{equation}
\tilde{g}_{ \tilde{\mu}\tilde{\nu}}= \left( \begin{array}{ccc}
g_{\mu\nu}-\phi A_\mu A_\nu & -\phi A_\mu  \\
-\phi A_\nu & -\phi  \end{array} \right) 
\end{equation}
Where $\mu$ and $\nu$  run over 0,1,2,3 and the tilde quantites run over 0,1,2,3,5. $A_{\mu}$ and $A_{\nu}$ are vector fields and $\phi$ is a scalar field. Now start with a source free spacetime (pure-gravity) in five dimensions. The corresponding action of the system is
\begin{equation}
S^5=-\int d^5x \sqrt{\tilde{g}}\tilde{R}
\end{equation}
\\
Where $\tilde{g}$ is the five dimensional determinant of the metric and $\tilde{R}$ is the five dimensional Ricci Scalar. After integrating out the extra dimension and performing a conformal rescaling of the metric $g_{\mu\nu}\rightarrow \sqrt{\phi}g_{\mu\nu}$, it is trivial to show that the four dimensional action can be written as:

\begin{equation}
S=\int d^4x \sqrt{g}(-R+\frac{1}{2} \partial_\mu \phi \partial^\nu-\frac{1}{4}e^{-\sqrt{3}\phi}F_{\mu\nu}F^{\mu\nu})
\end{equation}\\

From this action we retrieve the spin-2 graviton, the spin-1 photon and discover a spin-0 scalar field. This indeed was a success in unifying gravity with electromagnetism. However the unexpected existence of the scalar field was an embarrassment at the time. This stimulated later work on Brans-Dicke theories of gravity where the scalar field has the physical effect of changing the effective gravitational constant from place to place.
\\
$\indent$Kaluza's idea suffered from a very obvious drawback. If there is a fifth dimension, where is it? In 1926 Oskar Klein suggested that the fifth dimension compactifies so as to have the geometry of a circle of extremely small radius [18]. Thus, the space has topology $R^4\times S^1$. One way to envisage this spacetime is to imagine a hosepipe. From a long distance it looks like a one dimensional line but a closer inspection reveals that evey point on the line is in fact a circle.

\section{Extra Dimensions and the Casimir Effect}

$\indent$The Casimir effect is one of the most salient manifestations of the vacuum fluctuations. In its most elementary form, it is the interaction of a pair of neutral, parallel conducting planes whose existence modify the ground state of the quantum vacuum in the interior portion of the plates creating a force which attracts the plates to each other.  For the electromagnetic field, this force of attraction is:

\begin{equation}
F=-\frac{\pi A}{480}\frac{hc}{a^4}
\end{equation}\\

where A is the area of the plates, h is Planck's constant, c is the speed of light and a is the plate separation. For a review on the Casimir effect see [21].
\\
$\indent$The Casimir effect can be extended to regions of non-trivial topology. For example, on $S^1$, a circular manifold, one can associate $0$ and $2\pi$  with the location of the plates and the Casimir energy can be calculated. This becomes relevant when we consider models with additional spatial dimensions.
\\
$\indent$Two models with extra dimensions have become particularly popular in recent years. These are the Arkani-Hamed-Dimopoulos-Dvali (ADD) [3], [4] and the Randall-Sundrum (RS) models [22]. Both are attempts to explain what has become known as a heirachy problem in physics, which questions why the gravitational force is so much weaker than the other forces of nature. In the ADD model it is proposed that the force carriers of the standard model (the photon, $W^+$, $W^-$, $Z^0$ and the gluons) are constrained to exist on the usual four dimensional spacetime, which we will call the $\textit{bulk}$. Gravity however, is free to move both in the bulk and in the extra dimensions. These dimensions can be as large as $\mu m$ [17]. 
\\
$\indent$Because only gravity can propagate in the extra dimensions, we cannot `see' them as we see in the bulk through the electromagnetic field (light), nor can we observe them through weak or strong force interactions, which are also restricted to the bulk. It is the freedom of the graviton to propagate off of the bulk that dilutes the field strength accounting for the apparent weakness of gravity. In the RS models, it is the warping of the extra dimension that is the root cause of the weakness of gravity.
\\
$\indent$In both the ADD and RS models one can account for the cosmological constant by calculating the contribution of extra dimensional graviton fields to the Casimir energy and then associating this energy with the cosmological constant [11]. As an example we will demonstrate a calculation of the vacuum energy due to compactification in the ADD scenario in which the spacetime orbifold will be $M^4\times T^2$ . Matter fields will reside on the 3-brane (our universe) and gravity is free to propagate in the bulk. The calculations closely follow [15].
\\
$\indent$The four dimensional vacuum energy is given by:
\begin{equation}
\left<E_{vac}\right>=\sum_{n1,n2=1}^\infty \frac{1}{2}\int\frac{d^3k}{(2 \pi)^3}\sqrt{k^2+\frac{4\pi^2(n_1^2+n_2^2)}{R^2}} 
\end{equation}

Where R is the radius of the extra dimensions and $n_1$ and $n_2$ are the KK (Kaluza Klein) graviton modes. We will now perform dimensional regularization on the k integral by first rewriting it as:
\\
\begin{equation}
I(m_n)=\mu^{2\epsilon}\int \frac{d^3k}{(2 \pi)^3}({k^2+m_n^2})^{\frac{1}{2}-\epsilon} 
\end{equation}

Where we have made the substitution, $m_n^2=\frac{4\pi^2(n_1^2+n_2^2)}{R^2}$. The $\mu$ term is inserted for dimensional consistency and $\epsilon$ will be taken to zero. The integral can be expressed as a beta function which can be expanded to:

\begin{equation}
I(m_n)=-\frac{m_n^4}{32\pi^2}\left(\frac{1}{\epsilon}+2log2-\frac{1}{2}-log(\frac{m_n^2}{\mu^2})\right) 
\end{equation}

The vacuum energy can thus be expressed as:

\begin{equation}
\left<E_{vac}\right>=-\frac{1}{32\pi^2} \sum_{n1,n2=1}^\infty m_n^4\left(\frac{1}{\epsilon}+2log2-\frac{1}{2}-log(\frac{m_n^2}{\mu^2})\right)
\end{equation}

The summation is clearly divergent so we next use a zeta function regularization.

\begin{equation}
 \sum_{n1,n2=1}^\infty m_n^4=\left( \frac{4\pi}{R^2}\right) ^2\sum_{n1,n2=1}^\infty(n_1^4+n_2^4+n_1^2n_2^2)
\end{equation}

\begin{equation}
 =\left(\frac{4\pi}{R^2}\right)^2 2\left[\zeta(0) \zeta(-4)+\zeta^2(-2\right]
\end{equation}

Using the fact that $\zeta(-2m)=0$, for m any natural number, we obtain the vacuum energy:

\begin{equation}
\left<E_{vac}\right>_{4+2}=\frac{1}{32\pi^2} \sum_{n1,n2=1}^\infty m_n^4 log(\frac{m_n^2}{\mu^2})
\end{equation}

\begin{equation}
=\frac{\pi}{2R^4} \sum_{n1,n2=1}^\infty (n_1^2+n_2^2)^2 log(n_1^2+n_2^2)
\end{equation}

\begin{equation}
=-\frac{\pi^2}{R^4}\zeta(0)\zeta'(4)
\end{equation}

This result can be generalized to 4+n dimensions

\begin{equation}
\left< E_{vac}\right>=-\frac{\pi^2}{R^4} \left[ \frac{(2+n)(3+n)}{2}-1      \right]  \left[ \zeta(0) \right] ^{n-1}  \zeta'(4)
\end{equation}

where the graviton degrees of freedom is expressed in the brackets. This vacuum energy due to the massive KK modes is associated with the cosmological constant.

\section{Warp Drives}

$\indent$Numerous papers discussing the idea of warp drives have emerged in the literature in recent years. See for example [19]. The basic idea is to formulate a solution to Einstein's equations whereby a warp bubble is driven by a local expansion of spacetime behind the bubble and a contraction ahead of the bubble. One common feature of these papers is that their physical foundation is the General Theory of Relativity. An element missing from all the papers is that there is little or no suggestion as to $\textit{how}$ such a warp bubble may be created.
\\
$\indent$The aim of this paper is $\textbf{not}$ to discuss the plausibility of warp drive, the questions associated with violation of the null energy condition, or issues regarding causality. The aim of this paper is to suggest that a warp bubble could be generated using ideas and mathematics from $\textit{quantum field theory}$, and to hypothesize $\textit{how}$ such a bubble could be created by a sufficiently advanced technology.
\\
$\indent$By associating the cosmological constant with the Casimir Energy due to the Kaluza Klein modes of gravitons in higher dimensions, especially in the context of M-theory derived or inspired models, it is possible to form a relationship between $\Lambda$ and the radius of the compact extra dimension. We know from eqn.(16)

\begin{equation}
\left< E_{vac} \right> =\Lambda \propto \frac{1}{R^4}
\end{equation}

An easier way of developing this relationship is to put things in terms of Hubble's constant H which describes the rate of expansion of space per unit distance of space. 

\begin{equation}
H \propto \sqrt{\Lambda}, 
\end{equation}
or in terms of the radius of the extra dimension we have
\begin{equation}
H \propto \frac{1}{R^2}. 
\end{equation}

This result indicates that a sufficiently advanced technology with the ability to $\textit{locally}$ increase or decrease the radius of the extra dimension would be able to locally adjust the expansion and contraction of spacetime creating the hypothetical warp bubble discussed earlier. A spacecraft with the ability to create such a bubble will always move inside its own local light-cone. However the ship can utilize the expansion of space-time behind the ship to move away from some object at any desired speed or equivalently to contract the space-time in front of the ship to approach any object.
\\
$\indent$In the context of general relativity a similar phenomenology is produced for the case of anisotropic cosmological models, in which it is the $\textit{contraction}$ of the extra dimension that has the effect of expanding another [19]. For example consider a `toy' universe with one additional spatial dimension with the following metric

\begin{equation}
ds^2=dt^2-a^2(t)d\vec{x}^2-b^2(t)dy^2 
\end{equation}

In this toy universe we will assume spacetime is empty, that there is no cosmological constant and that all spatial dimensions are locally flat.

\begin{equation}
T_{\mu \nu}=\Lambda g_{\mu \nu}=0 
\end{equation}

The action of the Einstein theory of gravity generalized to five dimensions will be

\begin{equation}
S^{(5)}=\int d^4x dy \sqrt{-g^{(5)}}\left(\frac{M_5^2 }{16\pi }R^{(5)}\right) 
\end{equation}
\\
Solving the vacuum Einstein equations
\begin{equation}
G_{\mu\nu}=0,
\end{equation}

we obtain for the $G_{11}$ component

\begin{equation}
G_{11}=\frac{3\dot{a}(b\dot{a}+a\dot{b})}{a^2b}
\end{equation}

Rewriting $\frac{\dot{a}}{a}=H_a$ and $\frac{\dot{b}}{b}=H_b$ where $H_a$ and $H_b$ corresponds to the Hubble constant in three space and the Hubble constant in the extra dimension respectively, we find that solving for $G_{11}=0$ yields

\begin{equation}
H_a=-H_b.
\end{equation}

$\indent$This remarkable result indicates that in a vacuum, the shear of a contracting dimension is able to inflate the remaining dimensions. In other words the expansion of the 3-volume is associated with the contraction of the one-volume.  
\\
$\indent$ Even in the limit of flat spacetime with zero cosmological constant, general relativity shows that the physics of the $\textit{compactified}$ space effects the expansion rate of the non-compact space.The main difference to note here is that the quantum field theoretic result demonstrates that a $\textit{fixed}$ compactification radius can also result in expansion of the three-volume as is shown in eqn.(16) due to the Casimir effect, whereas the general relativistic approach suggests that a $\textit{changing}$ compactifification radius results in expansion.
Both add credibility to the warp drive concept presented here. 

\section{Energy Requirements}

$\indent$In this section we perform some elementary calculations to determine how much energy would be required to reach superluminal speeds. We also determine an absolute speed limit based on fundamental physical limitations.
\\
$\indent$The currently accepted value for the Hubble constant is 70 km/sec/Mpsc. A straightforward conversion into SI units gives $H = 2.17\times 10^{-18} (m/s)/m$. This tells us that one meter of space expands to two meters of space if one were prepared to wait two billion billion seconds or sixty five billion years. The fundamental idea behind the warp drive presented in this paper is to increase Hubbles constant such that space no longer expands at such a sedentary rate, but locally expands at an arbitrarily fast velocity. For example, if we want space to \textit{locally} expand at the speed of light, a simple calculation shows us by what factor we would need to increase H.

\begin{equation}
\frac{H_c}{H} \approx \frac{10^8}{10^{-18}}=10^{26}
\end{equation}

Where $H_c$ is the `modified' Hubble constant (subscript c for speed of light). This results implies that H would have to be increased by a factor of $10^{26}$ for space to expand at the speed of light. Since we know that $H \propto \frac{1}{R^2}$, we can naively form the relation

\begin{equation}
\frac{H_c}{H} = \frac{R^2}{R_c^2}=10^{26}
\end{equation}

or,

\begin{equation}
R_c=10^{-13}R
\end{equation}

Where $R_c$ is the modified radius of the extra dimension. This indicates that the extra dimensional radius must be $\textit{locally}$ reduced by a factor of $10^{13}$ to stimulate space to expand at the speed of light. In the ADD model the size of the extra dimension can be as large as $10^{-6}m$. If we use this number as a prototype extra-dimensional radius, this would have to be shrunk to $10^{-19}m$ for lightspeed expansion.
\\

$\indent$An interesting calculation is the energy required to create the necessary warp bubble. The accepted value of the cosmological constant is $\Lambda \approx 10^{-47} (GeV)^4$. Converting again into SI units gives $\Lambda \approx 10^{-10} J/m^3$ . Now, for a warp bubble expanding at the speed of light we would need to increase this again by a factor of $10^{52}$ as we have $H\propto \sqrt{\Lambda}$ . We can say

\begin{equation}
\Lambda_c=10^{52}\Lambda=10^{42}J/m^3
\end{equation}

where $\Lambda_c$ is the \textit{local} value of the cosmological constant when space is expanding at c. Let us consider a spacecraft of dimensions

\begin{equation}
V_{craft}=10m \times 10m \times 10m =1000m^3.
\end{equation}

If we postulate that the warp bubble must, at least, encompass the volume of the craft, the total amount of energy `injected' locally would equal

\begin{equation}
E_c=\Lambda_c \times V_{craft} =10^{45}J.
\end{equation}

Assuming some arbitrarily advanced civilization were able to create such an effect we might postulate that this civilization were able to utilize the most efficient method of energy production - matter antimatter annihilation. Using $E=mc^2$ this warp bubble would require around $10^{28} Kg$ of antimatter to generate, roughly the mass-energy of the planet Jupiter.
\\

$\indent$This energy requirement would drop dramatically if we assumed a thin-shell of modified spacetime instead of bubble encompassing the volume of the craft.

\section{Ultimate Speed Limit}

$\indent$It is known from string theory that the absolute minimum size for an extra dimension is the Planck length, $10^{-35}m$. This places an ultimate speed limit on the expansion of space based on the idea that there is a limit to the minimum radius of the extra dimension. From the above arguments it is straightforward to form the relation

\begin{equation}
\frac{H_{max}}{H}=\frac{R^2}{R_{min}^2}=\frac{10^{-12}}{10^{-70}}=10^{58}.
\end{equation}

Here $H_{max}$ is the maximum rate of spacetime expansion based on the minimum radius of the extra dimension $R_{min}$. In this formula we have again used a prototype extra-dimensional radius of $10^{-6}m$ which is the upper bound based on current experimental limits. Using these values and the known value of H in SI units we obtain

\begin{equation}
H_{max}=10^{58}H\approx 10^{40} (m/s)/m
\end{equation}

A quick conversion into multiples of the speed of light reveals

\begin{equation}
V_{max}=10^{32}c,
\end{equation}

which would require on the order of $10^{99}Kg$ of antimatter, more mass energy than is contained within the universe. At this velocity it would be possible to cross the known universe in a little over $10^{-15}$ seconds.
\\
$\indent$The authors wish to reiterate that the calculations in the previous two sections are extremely `back of the envelope' and merely serve as interesting figures to contemplate, based on the formula

\begin{equation}
\left< E_{vac} \right> = \Lambda \propto \frac{1}{R^4}.
\end{equation}

$\indent$Note that it is not really possible to travel faster than light in a local sense. One can however, make a round trip between two points in an arbitrarily short time as measured by an observer who remained at rest at the starting point. See [24] for details on violations of the null energy conditions and causality.

\section{Conclusions}

$\indent$In this paper we have calculated the vacuum energy due to extra dimensional graviton contributions to the Casimir energy and associated this energy with the cosmological constant. It has been shown that this energy is intimately related to the size of the extra dimension. We have picked the ADD model, however similar approaches can be used for alternative models, for example the RS model of warped extra dimension where a similar relation can be found.
\\
$\indent$ We have proposed that a sufficiently advanced civilization could utilize this relation to generate a localized expansion/contraction of spacetime creating a `warp bubble' in which to travel at arbitrarily high velocities. One vital aspect of future research would be $\textit{how}$ to locally manipulate an extra dimension. String theory suggests that dimensions are globally held compact by strings wrapping around them [6], [9]. If this is indeed the case, then it may be possible to even locally increase or decrease the string tension, or even locally counter the effects of some string winding modes. This would achieve the desired effect of changing the size of the extra dimensions which would lead to propulsion under this model. It would thus be prudent to research this area further and perform calculations as to the energies required to affect an extra dimension and to try and relate this energy to the acceleration a spacecraft might experience.

\section{Bibliography}

[1] Aghababaie, Y., Burgess, C., Parameswaran, S., Quevedo F., ``Towards a Naturally Small Cosmological Constant from Branes in 6-D Supergravity," Nucl.Phys. B680, 389-414, (2004).
\\

[2] Alcubierre, M., ``The Warp Drive: Hyperfast Travel Within General Relativity," Class.Quant.Grav.Letters 11, L73-L77, (1994).
\\

[3] Antoniadis,I, Arkani-Hamed, N., Dimopoulos, S., Dvali G., ``New dimensions at a millimeter to a Fermi and superstrings at a TeV". Phys. Lett. B 436: 257-263, (1998).
\\

[4] Arkani-Hamed, N., Dimopoulos, S., Dvali G., ``The Hierarchy Problem and New Dimensions at a Millimeter". Phys. Lett. B 436: 263-272, (1998).
\\

[5] Binetruy, P., Deffayet, C., Ellwanger, U., Langlois, D., ``Brane Cosmological Evolution in a Bulk with Cosmological Constant," Phys.Lett. B477, 285-291, (2000). 
\\

[6] Brandenberger, R., and Vafa, C., ``Superstrings in the Early Universe," Nucl. Phys. B136, 391 (1989)
\\

[7] Chen, P., and Gu, J., ``Casimir Effect in a Supersymmetry-breaking Brane World as Dark Energy", SLAC- PUB-10772, 2004.
\\

[8] Cherednikov, I., ``On Casimir Energy Contribution to Observable Value of the Cosmological Constant," Acta Phys. Polon. B33:1973-1977, (2002).
\\

[9] Cleaver, G and Rosenthal P., ``String Cosmology and the Dimension of Space-time," Nucl.Phys. B457, 621-642, (1995).
\\

[10] de Alwis, S., ``Brane World Scenarios and the Cosmological Constant," Nucl. Phys. B597, 263-278, (2001).
\\

[11] Ellwanger, U., ``Brane Universes and the Cosmological Constant," Mod.Phys.Lett. A20, 2521-2532, (2005).
\\

[12]  Garriga, J., Pujolas, O., Tanaka, T., ``Radion Effective Potential in the Brane World," Nucl.Phys. B605, 192-214,  (2001). 
\\

[13] Garnavich, P., ``Observational Evidence From Supernovae for an Accelerating Universe and a Cosmological Constant," Astron.J. 116:1009-1038, (1998).
\\

[14] Giudice, G., and   Rattazzi, R., "Theories with Gauge Mediated Supersymmetry Breaking," Phys. Rept. 322, 419-499, (1999).
\\

[15] Gupta, A., ``Contribution of Kaluza-Klein Modes to the Vacuum Energy in Models with Large Extra Dimensions and the Cosmological Constant", eprint arXiv:hep-th/0210069 
\\

[16] Kaluza, T., Sitz. Preuss. Akad. Wiss. Phys. Math. Kl 996 (1921).
\\

[17] Kapner, et al., ``Test of the Gravitational Inverse-Square Law below the Dark-Energy Length Scale," Phys. Rev. Lett. 98 021101 (2007).
\\

[18] Klein, O., Z. F. Physik 37 895 (1926); Klein, O.,Nature 118 516 (1926)
\\

[19] Levin, J., ``Inflation from Extra Dimensions," Phys.Lett. B343 69-75, (1995).
\\

[20] Lobo, S., Visser M., ``Fundamental limitations on 'warp drive' spacetimes," Class.Quant.Grav.21:5871-5892,2004.
\\

[21] Milton, K., ``The Casimir effect: Recent controversies and progress," J.Phys.A37:R209,2004
\\

[22] Morris, M., and Thorne, K,. and Yurtsever, U., ``Worm-holes, Time Machines, and the Weak Energy Conditions," Phys. Rev. Lett., 1446-1449, (1988).
\\

[23]  Randall, L., and Sundrum, R., ``An Alternative to Compactification," Phys. Rev. Lett. 83, 4690 (1999);
\\

[24] Tseytlin, A., and Vafa, C., ``Elements of String Cosmology," Nucl. Phys. B372, 443-466, (1992).
\\

[25] Visser, M., and Bassett, B., and Liberati, S., ``Perturbative superluminal censorship and the null energy condition," arXiv:gr-qc/9908023v1 (1999).

\end{document}